Title: The Mobius Band and the Mobius Foliation
Authors: I. P. Zois (American College of Greece, Deree College)
Comments: 15 pages
Report-no: ACG-DM-080101
Journal-ref:

This article presents some computations for a new topological invariant for foliations introduced some years ago
 by the author using techniques from noncommutative geometry, in particular the pairing between K-Theory and cyclic 
cohomology. The 
motivation came from theoretical physics, more specifically from flat vector bundles, in physics terminology from 
potentials which are pure gauge.

\documentstyle[11pt,amsfonts,amssymb,amsmath,amscd]{article}
\begin{document}
\title{The Mobius Band and the Mobius Foliation}
\author{Ioannis P. ZOIS
\\
School of Natural Sciences, Department of Mathematics
\\
The American College of Greece, Deree College
\\
 6 Gravias Sstreet, GR-153 42
\\
Aghia Paraskevi, Athens, Greece
\\
e-mail: izois@acg.gr}

\newtheorem{thm}{Theorem}
\newtheorem{defn}{Definition}
\newtheorem{prop}{Proposition}
\newtheorem{lem}{Lemma}
\newtheorem{cor}{Corollary}
\newtheorem{rem}{Remark}
\newtheorem{ex}{Example}

\newcommand{\Rat}{\mathbb Q}
\newcommand{\Real}{\mathbb R}
\newcommand{\RR}{\Real}
\newcommand{\Rh}{\hat{\Real}}
\newcommand{\Nat}{\mathbb N}
\newcommand{\Complex}{\mathbb C}
\newcommand{\HH}{\mathbb H_3}
\newcommand{\CC}{\Complex}
\newcommand{\Z}{\mathbb Z}

\newcommand{\Ea}{{\mathcal E}}
\newcommand{\Ta}{{\mathcal A}}
\newcommand{\Aa}{{\Ta_\infty}}
\newcommand{\Eb}{{C^*(D_1,D_2,X_2,\Omega)}}
\newcommand{\Tb}{{C^*(D_1,D_2,\hat\Omega)}}
\newcommand{\Ab}{{C^*(D_1,D_2,\Omega)}}
\newcommand{\aco}{\idl}
\newcommand{\aca}{\beta^{\|}}
\newcommand{\acb}{\beta^\perp}
\newcommand{\acc}{{\hat\tau}}

\newcommand{\Uu}{{\cal U}}
\newcommand{\Dd}{{\mathcal D}}
\newcommand{\Oo}{{\mathcal O}}
\renewcommand{\H}{{\mathcal H}}
\newcommand{\NN}{{\bf N}}
\newcommand{\ZZ}{{\bf Z}}
\newcommand{\Pp}{{\mathcal P}}
\newcommand{\Zz}{{\mathcal Z}}
\newcommand{\PP}{{\bf P}}
\newcommand{\LL}{\Lambda}
\newcommand{\LLL}{\Lambda\cup \infty}
\newcommand{\EE}{{\bf E}}
\newcommand{\Bb}{{\mathcal B}}
\newcommand{\Ww}{{\mathcal W}}
\newcommand{\Ss}{{\mathcal S}}

%Spuren
\newcommand{\Tr}{\mbox{\rm Tr}}  %Operatorspur
\newcommand{\TV}{{\mathcal T}}
\newcommand{\TVh}{\hat{{\mathcal T}}}
\newcommand{\tr}{\mbox{tr}}

\newcommand{\Rr}{{\mathcal R}}
\newcommand{\Nn}{{\mathcal N}}
\newcommand{\Cc}{{\mathcal C}}
\newcommand{\Jj}{{\mathcal J}}
\newcommand{\Ff}{{\mathcal F}}
\newcommand{\Ll}{{\mathcal L}}
\newcommand{\id}{{\mbox{\rm id}}}
\newcommand{\idl}{{\mbox{\rm\tiny id}}}
\newcommand{\eva}{{\mbox{\rm ev}}}
\newcommand{\eval}{{\mbox{\rm\tiny ev}}}

\def\essinf{\mathop{\rm ess\,inf}}
\def\esssup{\mathop{\rm ess\,sup}}

\newcommand{\bew}{{\bf Proof:}}
\newcommand{\eb}{\hfill $\Box$}

\newcommand{\x}{{\vec x}}
\newcommand{\y}{{\vec y}}
\newcommand{\n}{{\vec n}}
\renewcommand{\a}{{\vec a}}
\newcommand{\hull}{\Sigma}
\newcommand{\om}{\omega}
\newcommand{\oh}{{\hat{\om}}}

\newcommand{\Af}{{\Aa}_0}
\newcommand{\Tf}{{\Ta}_0}
\newcommand{\Ef}{{\Ea}_0}
\renewcommand{\H}{{\mathcal H}}

\newcommand{\bs}{\bigskip}
\newcommand{\ms}{\medskip}

\newcommand{\erz}[1]{\langle{#1}\rangle}
\newcommand{\pair}[2]{\erz{#1,#2}}
\newcommand{\diag}[2]{\mbox{\rm diag}(#1,#2)}
\newcommand{\tOmega}{{\Omega^s}}
\newcommand{\td}{{d^s}}
\newcommand{\tint}{{\int^s}}
\newcommand{\ttau}{{\tilde\tau}}
\newcommand{\talpha}{{\tilde\alpha}}
\newcommand{\tS}{{\tilde S}}
\newcommand{\ual}{{\underline{\alpha}}}
\newcommand{\hotimes}{{\hat\otimes}}
\newcommand{\K}{{\mathcal K}}
\newcommand{\Ypsilon}{{\Theta}}
\newcommand{\CA}{$C^*$-algebra}
\newcommand{\CF}{$C^*$-field}
\newcommand{\G}{\mathcal G}
\newcommand{\im}{\mbox{\rm im\,}}
\newcommand{\rk}{\mbox{\rm rk\,}}
\newcommand{\cp}{{\rtimes}}
\newcommand{\del}{{\bf \delta}}

\renewcommand{\a}{{\vec a}}
\newcommand{\at}{{\bf \tilde a}}
\newcommand{\cy}{\Phi}
\newcommand{\co}{\cy_\a}
\newcommand{\tco}{\cy_{\at}}

\newcommand{\ttimes}{{\tilde \rtimes}}
\renewcommand{\ss}{{\mathcal R}}
\renewcommand{\Cc}{{\mathcal C}}
\newcommand{\Ri}{(\RR\cup\infty)}
\newcommand{\Rd}{\hat\RR}
\newcommand{\dom}{\mbox{\rm dom}}
\newcommand{\fin}{\mbox{\rm\tiny fin}}

\newcommand{\enn}{\mu}
\newcommand{\ch}{\mbox{\rm ch}}
\newcommand{\supp}{\mbox{\rm supp}}
\newcommand{\chd}{\ch}
%%%%%%%%%%%%%%%%%%%%%%%%%%%%%
\newcommand{\Oh}{\hat\hull}

\newcommand{\hsp}{\RR^{d-1}\times\RR^{\leq 0}}
\newcommand{\IDS}{IDS}
\newcommand{\sv}[2]{\left(\begin{array}{c} #1 \\ #2 \end{array}\right)}

\bibliographystyle{amsalpha}

\maketitle

\begin{abstract}
Some years ago we introduced a new topological invariant for foliated manifolds using techniques from noncommutative 
geometry, in particular the pairing between K-Theory and cyclic cohomology. The motivation came from flat principal 
$G$-bundles where the base space is a non simply connected manifold. The computation of this invariant is quite complicated. 
In this article we try to perform certain computations for the Mobius band (or Mobius foliation) 
which is an
interesting  nontrivial example of foliations; this example has a key feature: it is the simplest case
of a large class of examples of foliations, that of bundles with
discrete structure groups which also includes the foliations given by flat
vector (or $G$-principal) bundles. We shall see that the Mobius foliation example
also helps one to understand another large class of examples of foliations
coming from group actions on manifolds which are not free.


\end{abstract}

\section{Introduction}

In this article we study the Mobious band in an attempt to perform a
non-trivial computation for an operator algebraic invariant for foliations
introduced in \cite{z}. The construction of this invariant is quite complicated, it involves various stages:
first one has to determine the holonomy groupoid of the foliation, then find its corresponding $C^{*}$-algebra, then 
compute its K-Theory and its cyclic cohomology and construct natural classes in both and finally apply Connes' pairing 
between 
K-Theory an cyclic cohomology. In this preliminary version of the final article, we shall present some of these steps 
for the example of the Mobius 
band (or the Mobius foliation) along with some questions. But before that we shall recall some basic facts about foliations
 (throughout this work all manifolds are assumed to be smooth).\\

Let $M$ be a smooth
connected and closed manifold of dimension $m$. A codim-$q$
(and hence of dimension $(m-q)$) foliation on $M$ is given by an
\emph{integrable} subbundle $V$ of the tangent
bundle $TM$ of $M$ where the dimension of the fibre of $V$ is $(m-q)$. Quite
often $V$ is called the tangent bundle of the foliation as opposed to the
quotient bundle $TM/V$ which is called the \textsl{transverse bundle} of the
foliation. The effect is that given a $V$ as above,
$M$ can be written as the \emph{disjoint union} of the leaves of the foliation
which are \textsl{immersed}, connected submanifolds of $M$, all of
 the same dimension equal to $(m-q)$. The topology of the leaves may vary
drastically: some may be compact but others not and more importantly their
fundamental groups are different. From these two differences one can see that
foliations are important generalisations of the total space of fibre
bundles because in a fibre bundle the total space is the disjoint union of the
fibres which are essentially the same manifold (ie homoeomorphic or
diffeomorphic) to a fixed model manifold called typical fibre.\\

If the foliation has a transverse bundle which can be oriented, an equivalent
local definition of a codim-$q$ foliation is given by a
nonsingular decomposable $q$-form
$\omega$ satisfying the integrability condition $\omega\wedge d\omega =0$.
The leaves are the submanifolds whose tangent vectors vanish on $\omega$.
By the Frobenius theorem the set of smooth sections of $V$ denoted
$C^{\infty} (V)$
 form a Lie subalgebra of the Lie algebra $C^{\infty} (TM)$ of vector fields of
 $M$ (seen as sections of its tangent bundle). Dually, the annihilator ideal
$I(V)$ of $V$ consisting of differential
forms vanishing on the leaves (ie on sections of $V$) is closed under the de
Rham differential $d$, namely since the annihilator ideal is a graded ideal we
write
$d(I^{*}(V)\subseteq I^{*+1}(V)$. In the codim-1 case one
 can show that this annihilator ideal of $V$ can be generated by $\omega$
itself and thus the integrability relation
$\omega\wedge d\omega =0$ is equivalent
 to $d\omega =\omega\wedge\theta$ where $\theta$ is another 1-form which has
the property that it is \emph{closed when restricted on every leaf}, thus
defining a class in the first cohomology group of every leaf. This 1-form
$\theta$ is sometimes called the (partial) flat Bott connection on the
transverse bundle of the foliation. Moreover for any codimension
$d\theta \in I^{2}(V)$. The form
$\omega$ can be multiplied with a
nowhere vanishing function $f$ without changing the foliation. The effect it
will have on $\theta$ is that we add an exact form. Thus the cohomology class
that $\theta$ defines on every leaf does not change. The Godbillon-Vey class of
 the foliation $V$ is the real $(2q+1)$ de Rham cohomology class
$\theta\wedge (d\theta)^{q}$ and it does not depend on the coices of $\omega$
and $\theta$, it only depends on $V$. Its existence follows from Bott's
vanishing theorem: if a codim-$q$ subbundle $V$ of $TM$ is integrable, then the
Pontrjagin classes of the transverse bundle $TM/V$ vanish in degrees greater
than twise the codimension. \emph{The holonomy of the flat Bott connection on
every leaf is the \textsl{infinitesimal part} of the germinal holonomy of that
particular leaf}; to quantify this information about the infinitesimal germinal
 holonomy
 of a leaf in the codim-1 case one can either take the trace or the determinant
 composed with the logarithm; in the later case one obtains the cohomology
class that $\theta$ defines in the 1st de Rham cohomology group of the specific
 leaf (see \cite{cc} page 66). Moreover $\theta$ also
defines a 1st tangential cohomology class. But
one has to realise that $\theta$ is closed only when restricted to a particular
 leaf and it carries the infinitesimal information of the holonomy of that
particular leaf. The GV-class however is a real cohomology class of $M$ which
carries information about the infinitesimal holonomy of the \emph{foliation as
a whole} (ie somehow the GV-class is an average over all leaves of the
det composed with log of the infinitesimal germinal holonomy of each one of
them and the tfcc is an average over all leaves of the traces of the
infinitesimal germinal holonomy since cyclic cohomology contains traces;
remaining is to see what is the average over all leaves of the Ray-Singer
analytic torsion). Duminy's
theorem says that in the codim-1 case only the resilient leaves contribute to
the GV-class.\\

The simplest example of foliated manifolds is the Cartesian product of
 two manifolds $M\times N$.\\

The second example is submersions: let $P$ and $M$ be smooth manifolds of
dimension $p$ and $m\leq p$ respectively and let $f:P\rightarrow M$ be a
submersion, namely $rank(df)=m$. By the implicit function theorem $f$ induces a
codim-$m$ foliation on $P$ where the leaves are the components of $f^{-1}(x)$
for $x\in M$. \emph{Fibre bundles} are particular examples of this sort.\\

The third nontrivial example of foliations is given by \emph{bundles with
discrete structure group}. Let $\pi :P\rightarrow M$ be a differentiable fibre
bundle with fibre $F$ and $p=dim(P)$, $q=dim(F)$ and $m=dim(M)$; clearly
$p=q+m$. A bundle is defined by an open covering $\{U_{a}\}_{a\in
 A}$ of $M$, diffeomorphisms $h_{a}:\pi ^{-1}(U_{a})\rightarrow U_{a}\times F$
called local trivialisations and transition functions 
$g_{ab}:U_{a}\cap U_{b}\rightarrow Diff(F)$ such that $h_{a}\circ h_{b}^{-1}(x,y)=(x,g_{ab}(x)(y))$ and
moreover the transition functions satisfy the cocycle relation in triple
intersections $g_{ab}\circ g_{bc}=g_{ac}$. If the transition functions are
\textsl{locally constant}, then the bundle is said to have discrete structure
group. Under this assumption the codim-$q$ foliations of $\pi ^{-1}(U_{a})$
given by the submersions fit nicely together to give a foliation on
$P$. \emph{Flat} vector or principal $G$-bundles with $G$ a compact and
connected Lie group are of this sort, namely
vector bundles or principal $G$-bundles equipped with a flat connection
(a connection with zero curvature).

Every such bundle can be constructed in the following way: let $\phi :
\pi _{1}(M)\rightarrow Diff(F)$ be a group homomorphism and we denote by $G$
the image of $\pi _{1}(M)$ into $Diff(F)$ via $\phi $. Moreover let
$\tilde{M}$
denote the universal covering space of the base manifold $M$. Then
$\pi _{1}(M)$ acts jointly on the product space $\tilde{M}\times F$ as follows:
 it acts via deck transformations on $\tilde{M}$ and by $\phi $ on $F$. So we
can define $P:=\tilde{M}\times _{\pi _{1}(M)}F$. This action is free and
properly discontinuous, hence $P$ is a foliated manifold of codim-$q$ (and
hence of dimension $m$) where
$q=dim(F)$. The leaves look like many valued cross-sections of the bundle
$\pi :P\rightarrow M$ and in fact $\pi$ restricted to any leaf is a covering
map. To see this, note that if $L_{x}$ is the leaf through the point
corresponding to
$\tilde{M}\times \{x\}\subset\tilde{M}\times F$, then $L_{x}$ is diffeomorphic
to $\tilde{M}/G_{x}$ where $G_{x}=\{g\in \pi _{1}(M):\phi (g)(x)=x\}$ denotes
 the isotropy group at $x$.\\

The simplest case of a bundle with discrete structure group is the
\emph{Mobius band}. We shall focus on this example here. We take the base space
$M$ to be $S^{1}$ with local coordinate $s$ and we know that
$\pi _{1}(S^{1})={\bf Z}$, and the universal
 covering space of $S^{1}$ is ${\bf R}$. Then ${\bf Z}$ acts on ${\bf R}$ via
deck transformations $s\mapsto s+1$  The fibre $F$
will be ${\bf R}$ with local coordinate denoted $r$. Then we pick
$\phi \in Diff({\bf R})$ to be given by
$\phi (r)=-r$ for $r\in {\bf R}$. Then the product space
${\bf R}\times {\bf R}$ has a ${\bf Z}$ action given by
$(s,r)\mapsto (s+1,-r)$. The quotient space by this ${\bf Z}$ action is the
\emph{Mobius band} $M:={\bf R}\times _{\bf Z}{\bf R}$. Each leaf  $L_{r}$ is a
circle wrapping around \emph{twice} except for the core circle (corresponding
to $r=0$) which wraps around only once.\\

What about our local definition of foliations? Well, for the Mobius band we
have that its transverse bundle is not orientable so there does not exist a
definition involving 1-forms.\\

An important remark now: the Mobius band $M$ can be considered in two ways:
either as the total space of a vector bundle over $S^{1}$ with fibre
${\bf R}$.
 In this case if we squeeze every fibre to a point we get of course as a result
 $S^{1}$ as the quotient space and
$K^{i}(S^{1})={\bf Z}$ with $i=0,1$. However if we consider the
Mobius band as a foliated manifold as
above and we squeeze every leaf to a point we get ${\bf R}_{r\geq 0}$ as the
quotient space and we know that $K^{i}({\bf R}_{r\geq 0})=0$ for $i=0,1$ (here
since the space is only locally compact we have to use K-Theory with compact
supports).

Next we want to compute the \emph{holonomy groupoid} of the Mobius band but
before doing that let us briefly recall the key notion of holonomy for
foliations and how these data can be organised to what is called the germinal
holonomy groupoid of the foliation.

The holonomy essentially tells us how leaves assemble together to give the
foliation and it is the most important intrinsing notion for foliations. It
encodes information about the fundamental groups of the leaves (which as we
emphasised
above they can vary enormously from leaf to leaf in sharp contrast to the
fibres in a fibre bundle which are the same manifold) along with information
about how the leaves fit nicely together in order to have the manifold as their
 disjoint union. The key difference with fibre bundles here is the spiralling
phenomenon: leaves may spiral repeatedly over each other without intersecting.

 Let $x$ be a point on a foliated manifold say $M$ and let $L_{x}$ denote the
unique leaf trough the point $x$. Then the holonomy group $G_{x}^{x}$ over the
point $x$ is defined to be the image of the surjective homomorphism
$h_{x}: \pi _{1}(L_{x})\rightarrow Diff(F,x)$ where $F$ is a transversal and
$Diff(F,x)$ denotes the germs of local diffeomorphisms at $x$ which fix $x$.
This map $h$ is precisely the \emph{holonomy} of the foliation which gives
diffeomorphisms between transversals following the plaque to plaque process
using regular foliated atlases (``sliding transversals along leaves'').
 Note the similarities with the holonomy of a connection on a vector bundle:
the fibres are the transversals and the plaque to plaque process corresponds to
 parallel transport of vectors using the connection (for more details see
for example \cite{cc}).\\

Now one can simply consider the disjoint union of all the holonomy groups
$G^{x}_{x}$ for all points $x$ of the foliated manifold say $M$ and this forms
a groupoid called the \emph{germinal holonomy groupoid} of the foliation.
A groupoid
can be defined as a small category with inverses and the space of objects is
the foliated manifold $M$ itself (for more details see \cite{connes} or
\cite{moore}). Perhaps a more concise notation is the following: the holonomy
groupoid of a foliation $V$ on a manifold $M$ as a set consists of triples
$G:=\{(x,h,y): x\in M, y\in L_{x}, h$ the holonomy class of a path (usually a
 loop) from $x$ to $y\}$. \\

We now turn to the case of a bundle with discrete structure group $P$, with
fibre $F$ and base smanifold $M$ with discrete
structure group given by  the group homomorphism
$\phi :\pi _{1}(M)\rightarrow Diff(F)$ as described above; the total space is
defined via
$P:=\tilde{M}\times _{\pi _{1}(M)}F$ where $\pi _{1}(M)$ acts via \emph{deck
transformations on} $\tilde{M}$ and \emph{by} $\phi $ \emph{on} $F$. Let $G$
denote the
image of $\pi _{1}(M)$ into $Diff(F)$ under $\phi$ and for each $x\in F$ let
$G_{x}:=\{g \in G :gx=x\}$ denote the isotropy group at
$x$ and let $G^{x}:=\{g \in G :gy=y$ for all $y$ in some
neighborhood of $x$ in $F$\} denote the stable isotropy group at $x$. The leaf
$L_{x}$ through $x$ (which is the image of $\tilde{M}\times \{x\}$ in $P$)
can be expressed as $L_{x}=\tilde{M}/G_{x}$ where $G_{x}$ acts on
$\tilde{M}$ by deck transformation.\\

 Then $G^{x}$ is a normal subgroup of $G_{x}$ and the holonomy
group
$G_{x}^{x}$ of the foliation over $x$ is simply
$G_{x}^{x}=G_{x}/G^{x}$. Let now $m\in M$ be a basepoint and
$\tilde{m}\in\tilde{M}$ be a preimage of $m$ and let $N$ be the image of
$\tilde{m}\times F$ in $P$. The map $\tilde{m}\times F\rightarrow N$ is a
diffeomorphism since $\pi _{1}(M)$ acts freely on $\tilde{M}$, so $N$ is a
copy of the fibre $F$ sitting as a complete transversal in the foliated
manifold $P$. Thus we see that \textsl{all bundles with discrete structure
group admit a complete transversal}. This is important because if a foliation
 admits a complete transversal say $N$, both its holonomy groupoid and its
$C^{*}$-algebra completion simplify drastically by the Hilsum-Skandalis
theorem: namely the holonomy groupoid reduces to $G_{N}^{N}$ which is
simply the restriction of the full holonomy groupoid over the complete
transversal $N$ (if we see
the holonomy groupoid of the foliation as a small category with inverses with
objects $P$, then $G_{N}^{N}$ is a full subcategory with objects $N$)
and the $C^{*}$-completion of the holonomy groupoid is Morita Equivalent to the
$C^{*}$-completion of just $G_{N}^{N}$. Hence all we have to understand is
$G_{N}^{N}$. In fact $G_{N}^{N}$ is completely determined by the action of
$G$ on $F$ (remember that $G$ is the image of $\pi _{1}(M)$ into
$Diff(F)$ under $\phi$). More precisely one has the following homoeomorphism of
topological groupoids:
$$G_{N}^{N}\cong (F\times G)/\sim $$
where the equivalence relation is given by $(x,\gamma )\sim (y,\delta )$ if and
 only if $x=y$ and $\delta ^{-1}\gamma $ lies in the stable isotropy group
$G^{x}$. Perhaps a better way to rewrite the above would be that
$G^{N}_{N}=F\rtimes _{\phi}G$ and thus it is clear that the corresponding
$C^{*}$-algebra to this foliation will be Morita equivalent to
$C_{0}(F)\rtimes _{\phi}G$. If we have the particular case of a
\emph{flat} principal $H$-bundle over $M$ where $H$ is the structure Lie
group, then the \emph{holonomy} of the flat connection defines a map
$h:\pi _{1}(M)\rightarrow H$ and clearly in our discussion above $G$ will be
the homomorphic image of the fundamental group onto $H$ via $h$, namely
$G=h(\pi _{1}(M))\subset H$ and hence the
corresponding $C^{*}$-algebra to the foliation will be Morita equivalent to
$C_{0}(H)\rtimes _{h}G$. Would like to see what groups can appear as
holonomy groups of flat connections and if the action of the holonomy $h$ they
has fixed points (well, it can have as we see from teh Mobius foliation below),
since both these issues are important in order to compute the $K_{0}$ group of
the corresponding crossed product algebra.\\

Now for the Mobius band $M:={\bf R}\times _{\bf Z}{\bf R}$ foliated by circles,
these circles correspond to the images of ${\bf R}\times \{r\}$ for various
values of $r$: if $r\neq 0$ then $\pi _{1} (L_{r})=\pi _{1}(S^{1})={\bf Z}$
acts trivially on $Diff({\bf R},r)$ and hence $G^{r}_{r}=0$ (note these circles
 wrap arround twice before they close). However the holonomy group $G_{0}^{0}$
of the middle circle which wraps around only once is the group ${\bf Z}_{2}$
since the diffeomorphism  $\phi (r)=-r$ which creates $M$ does lie in
$Diff({\bf R},0)$ and $\phi ^{2}=1$. Thus the group $G={\bf Z}_{2}$
\emph{acts non-freely} since $0$ is a fixed point. However this fixed
point is isolated with no interior and thus we have that
$G_{N}^{N}={\bf R}\times {\bf Z}_{2}$ and this is the holonomy groupoid of the
Mobius foliation as a topological space.
Now ${\bf Z}_{2}=\{\pm 1\}$ but we denote these two elements as $e=1$ for the
identity element and $\epsilon =-1$ the other one.
If we want to take the multiplication
into account as well, this will be $\Gamma ={\bf R}\rtimes _{\phi}{\bf Z}_{2}$
where the action $\phi $ of ${\bf Z}_{2}$ onto ${\bf R}$ is given by
``flipping the sign''. Its $C^{*}$-algebra completion is then
$A:=C_{0}({\bf R})\rtimes _{\phi}{\bf Z}_{2}$ where $C_{0}({\bf R})$ denotes
continuous complex valued functions defined on ${\bf R}$ which vanish at $0$
and infinity. Let us recall that as a linear space
$C_{0}({\bf R})\rtimes _{\phi}{\bf Z}_{2}$ consists of continuous maps
$F:{\bf Z}_{2}\rightarrow C_{0}({\bf R})$.
The product in $C_{0}({\bf R})\rtimes _{\phi}{\bf Z}_{2}$ is given by
$(F_{0}*F_{1})(\xi )\sum _{n\in {\bf Z}/2}F_{0}(n)\phi (n)F_{1}(n^{-1}\xi )$
where $n,\xi\in {\bf Z}_{2}$, and $\phi (e)=e$ for the identity element whereas
$\phi (\epsilon)(F)(x)=F(-x)$.\\

Let us be a little more explicit: since $M$ locally looks like
$S^{1}\times {\bf R}$, we choose local coordinates $(s,r)$ as above. Then if
$\pi :M\rightarrow S^{1}$ is the bundle projection, we pick as a complete
transversal $N$ the space $\pi ^{-1}(0)$ which is just a copy of ${\bf R}$.
Then the holonomy groupoid $G$ of the Mobius foliation according to what we
mentioned above for arbitrary bundles with discrete structure group is
homoeomorphic to simply $G^{N}_{N}$ where $N$ is a complete transversal.\\

The next order of bussiness is to compute the K-Theory of the groupoid
$C^{*}$-algebra completion. This is complicated because this algebra is
nonunital and hence we have to attach a unit and then throw it away.
We shall use the fact that in general any short exact sequence of algebras
$$0\rightarrow J\rightarrow E\rightarrow B:=E/J\rightarrow 0$$
gives rise to a 6-term long exact sequence in K-Theory:
\begin{equation}
\begin{CD}
K_{0}(J)@>>>K_{0}(E)@>>>K_{0}(E/J)\\
@VVV @VVV     @VVexpV\\
K_{1}(E/J)@<<<K_{1}(E)@<<<K_{1}(J)\\
\end{CD}
\end{equation}
 We shall apply this
 in oredr to compute the 0th K-group of the algebra
$C_{0}({\bf R})\rtimes _{\phi}{\bf Z}_{2}$ corresponding to the Mobius
foliation.\\

Remember that the group ${\bf Z}_{2}$ action on ${\bf R}$ has $0$ as a fixed
point; consider the map $ev_{0}:C_{0}({\bf R})\rightarrow {\bf C}$ which is
given by evaluating functions at the (fixed point) zero. Then one has the
following short exact sequence $(1)$:
$$0\rightarrow C_{0}({\bf R}^{-})\oplus C_{0}({\bf R}^{+})\hookrightarrow C_{0}({\bf R})\rightarrow {\bf C}\rightarrow 0$$
where ${\bf R}^{-}=(-\infty ,0)$, ${\bf R}^{+}=(0,\infty )$, $C_{0}({\bf R}^{+})$ denotes continuous functions vanishing both at $0$ and $+\infty $ (and
similarly for the $-$ sign).\\

Clearly since
${\bf R}^{-}\cong {\bf R}^{+}\cong {\bf R}$ are homoeomorphic, then
$C_{0}({\bf R}^{-})\oplus C_{0}({\bf R}^{+})\cong C_{0}({\bf R})\oplus C_{0}({\bf R})$. Using the
following well-known results that $K_{0}({\bf C})={\bf Z}$, $K_{1}({\bf C})=0$,
$K_{0}(C_{0}({\bf R}))=0$ and $K_{1}(C_{0}({\bf R}))={\bf Z}$ along with the
additivity property of the K-functor we get the
following corresponding K-Theory 6-term l.e.s.:

\begin{equation}
\begin{CD}
0@>>>0@>>>{\ZZ}\\
@VVV @VVV     @VVexpV\\
0@<<<{\ZZ}@<<<{\ZZ}^{2}\\
\end{CD}
\end{equation}

Since $0$ is a fixed point we can readily incorporate the ${\bf Z}_{2}$ action
onto the first s.e.s. and we directly get the second s.e.s (namely that the map
 evaluation at point $0$ is compatible with the ${\bf Z}_{2}$-action):
$$0\rightarrow (C_{0}({\bf R}^{-})\oplus C_{0}({\bf R}^{+}))\rtimes _{\phi}{\bf Z}_{2}\hookrightarrow C_{0}({\bf R})\rtimes _{\phi}{\bf Z}_{2}\rightarrow {\bf C}\rtimes _{\phi}{\bf Z}_{2}\rightarrow 0$$

We know that ${\bf C}\rtimes _{\phi}{\bf Z}_{2}$ is isomorphic to
${\bf C}\oplus {\bf C}$ and hence their K-groups are equal. From what we
mentioned above about the K-groups of ${\bf C}$ and the additivity of the
K-functor we can hence deduce that
$K_{0}({\bf C}\rtimes _{\phi}{\bf Z}_{2})={\bf Z}^{2}={\bf Z}\oplus {\bf Z}$
and that $K_{1}({\bf C}\rtimes _{\phi}{\bf Z}_{2})=0$. Moreover
$(C_{0}({\bf R}^{-})\oplus C_{0}({\bf R}^{+}))\rtimes _{\phi}{\bf Z}_{2}$ is
isomorphic to $M_{2}(C_{0}({\bf R}))$, the isomorphism being denoted $\Psi$ which takes an $F\in C_{0}({\bf R}^{-})\oplus C_{0}({\bf R}^{+})\rtimes _{\phi}{\bf Z}_{2}$ and it is mapped to\\

$$\Psi (F)=\sv{\psi ^{-}(F_{1}(e))\quad \psi ^{-}(F_{1}(\epsilon ))}{\psi ^{+}(F_{2}(\epsilon ))\quad \psi ^{+}(F_{2}(e))}$$\\

where $\psi^{\pm}:C_{0}({\bf R}^{\pm})\cong C_{0}({\bf R})$ with $\psi ^{+}(f)=f\circ exp$ and $\psi ^{-}(f)=f\circ (-exp)$.\\

Yet  $M_{2}(C_{0}({\bf R}))$ is Morita Equivalent to
$C_{0}({\bf R})$ and thus they have the same K-groups, so we get that
$K_{0}((C_{0}({\bf R}^{-})\oplus C_{0}({\bf R}^{+}))\rtimes _{\phi}{\bf Z}_{2})=0$ and that $K_{1}((C_{0}({\bf R}^{-})\oplus C_{0}({\bf R}^{+}))\rtimes _{\phi}{\bf Z}_{2})={\bf Z}$.
 Then we apply the corresponding 6-term K-Theory l.e.s.
to the second s.e.s. which incorporates the ${\bf Z}_{2}$-action and we get:

\begin{equation}
\begin{CD}
0@>>>K_{0}(C_{0}({\bf R})\rtimes _{\phi}{\bf Z}_{2})@>>>{\bf Z}\oplus {\bf Z}\\
@VVV @VVV     @VVexpV\\
0@<<<K_{1}(C_{0}({\bf R})\rtimes _{\phi}{\bf Z}_{2})@<<<{\bf Z}\\
\end{CD}
\end{equation}

which gives the result that\\

$K_{0}(C_{0}({\bf R})\rtimes _{\phi}{\bf Z}_{2})=Ker(exp)$\\

and\\

$K_{1}(C_{0}({\bf R})\rtimes _{\phi}{\bf Z}_{2})=Im(exp)$\\

In order to try to compute the groups explicitly we need more work and by
performing the computation we shall also determine the generators of the
groups as well. Let us start by recalling some known facts:
If $A$ is an associative algebra, $p\in A$ is called a projection if
$p^{2}=p$ with $0$ being the trivial projection; for two projections $p$, $q$
we write $p<q$ if $pq=p$. A projection is called minimal if we cannot find a
smaller one. If $A$ is unital we can easily construct projections in
$M_{n}(A)$ (the algebra of $n\times n$ matrices with entries from $A$) just by
considering the diagonal matices with the unit in the diagonal and each one
will corresppond to the free module over $A$ of rank $n$.

Let
$$ p_{+} =\frac{1}{2} \sv{1\quad 1}{1\quad 1}$$

and

$$ p_{-} =\frac{1}{2} \sv{1\quad -1}{-1\quad 1}$$\\

denote minimal projections in ${\bf C}\oplus {\bf C}$ (these can be used also
as generators of the algebra ${\bf C}\oplus {\bf C}$) which under the
isomorphism correspond to minimal projections $\frac{1}{2}(1_{e}+1_{\epsilon})$
 and $\frac{1}{2}(1_{e}-1_{\epsilon})$ in ${\bf C}\rtimes _{\phi}{\bf Z}_{2}$
where evidently $1_{e}:{\bf Z}_{2}\rightarrow {\bf C}$ denotes the function
giving $1$ on the identity
element i.e. $1_{e}(e)=1$ and $1_{e}(\epsilon)=0$ and similarly for
$1_{\epsilon}$.
Their corresponding K-classes wil be denoted $[p_{+}]$ and $[p_{-}]$ and these
are the two generators of
$K_{0}({\bf C}\rtimes _{\phi}{\bf Z}_{2})={\bf Z}^{2}$, hence any element in
the 0th K-group can be written as a finite integer linear combination of these
two elements. We would like to see under the exponential map
$exp: K_{0}({\bf C}\rtimes _{\phi}{\bf Z}_{2})(={\bf Z}^{2})\rightarrow K_{1}((C_{0}({\bf R}^{-})\oplus C_{0}({\bf R}^{+}))\rtimes _{\phi}{\bf Z}_{2})(={\bf Z})$what happens to the generators.\\

So we want to calculate $exp([p_{\pm}])=[e^{2i\pi\chi\frac{1}{2}(1_{e}\pm 1_{\epsilon})}]=[u_{\pm}]$, where $\chi :{\bf R}\rightarrow {\bf C}$ and $\chi 1_{e}:{\bf Z}_{2}\rightarrow C_{0}({\bf R})$ such that
$\chi 1_{e}(e)=\chi$ and $\chi 1_{e}(\epsilon)=0$ and moreover
$ev_{0}\chi 1_{e}=\chi (0)1_{e}=1_{e}$, $\chi (-\infty)=\chi (\infty)=0$ and
$\chi (0)=1$ and similarly for $\chi 1_{\epsilon}$. Note here that
$\chi\frac{1}{2}(1_{e}\pm 1_{\epsilon})$ denotes the \textsl{lifting} of the
projections $p_{\pm}=\frac{1}{2}(1_{e}\pm 1_{\epsilon})$ originally in
${\bf C}\rtimes _{\phi}{\bf Z}_{2}$ to self-adjoint elements in
$C_{0}({\bf R})\rtimes _{\phi}{\bf Z}_{2}$.\\

More concretely one can understand $u:{\bf Z}_{2}\times {\bf R}\rightarrow {\bf C}$ thinking it as a function from ${\bf Z}_{2}\rightarrow C_{0}({\bf R})$ with
$u_{\pm}(-\infty)=u_{\pm}(\infty)=1=u_{\pm}(0)$.\\

Hence we have to calculate $exp([1_{e}])=[e^{2i\pi\chi 1_{e}}]$. Now we said
that
for any associative algebra $A$, this is Morita equivalent to $M_{2}(A)$ hence
$K_{1}(M_{2}(A))\cong K_{1}(A)$ and if $[u]$ is in $K_{1}(A)$ the
corresponding element in  $K_{1}(M_{2}(A))$ is:

$$[\sv{u\quad 0}{0\quad 1}].$$

We mention the well-known fact that the following elements are equal
(as K-classes, hence homotopic as projections):

$$[\sv{u\quad 0}{0\quad 1}]=[\sv{1\quad 0}{0\quad u}],$$

 which we shall use later on. We pick a function
$\theta :{\bf R}\rightarrow {\bf C}$
which is $0$ at $-\infty$ and $1$ at $\infty$ so $e^{2i\pi\theta}\in C_{0}({\bf R})$. Thus using the explicit isomorphism $\Psi$ between
$C_{0}({\bf R}^{-})\oplus C_{0}({\bf R}^{+})\rtimes _{\phi}{\bf Z}_{2}$ and
 $M_{2}(C_{0}({\bf R}))$ plus the relation between the generators of the
$K_{1}$'s of the Morita equivalent algebras $A$ and $M_{2}(A)$ we deduce that
the generator of
$K_{1}((C_{0}({\bf R}^{-})\oplus C_{0}({\bf R}^{+}))\rtimes _{\phi}{\bf Z}_{2})$
is:

$$[\sv{(\psi^{-})^{-1}(e^{2i\pi\theta})\quad 0}{0\quad 1}]=[\sv{1\quad 0}{0\quad (\psi^{+})^{-1}(e^{2i\pi\theta})}].$$

Hence we conclude that (using also the relation between $\theta$ and $\chi$)

$$exp([1_{e}])=[\sv{(\psi^{-})^{-1}(e^{2i\pi\theta})\quad 0}{0\quad 1}]+
[\sv{1\quad 0}{0\quad (\psi^{+})^{-1}(e^{2i\pi\theta})}]=2,$$

 namely $exp([p_{+}]+[p_{-}])=2[e^{2i\pi\theta}]$, so the sum of the generetors
 of $K_{0}({\bf C}\rtimes _{\phi}{\bf Z}_{2})$ equals $2$ and hence
$[p_{+}]+[p_{-}]$ is NOT in $Ker(exp)$. So we still want to find the kernel of
the exponential map.\\

We know that $ev_{0}\chi p_{\pm}=p_{\pm}$ then $exp[p_{\pm}]=[exp(2i\pi\frac{1}{2}(1_{e}\pm 1_{\epsilon})\chi )]$ but $exp(2i\pi\frac{1}{2}(1_{e}\pm 1_{\epsilon})\chi )=[e^{2i\pi\chi}\frac{1}{2}(1_{e}\pm 1_{\epsilon})]+\frac{1}{2}(1_{e}\pm 1_{\epsilon})^{\perp}$ (which follows from the general identity for a
projection $p$ commuting with $f$ that $e^{2ipf\pi}=e^{2if\pi}p+(1-p)$). Hence
$(e^{2i\pi\chi}-1)\in C_{0}({\bf R}^{-})\oplus C_{0}({\bf R}^{+})$ because it
vanishes at $0$ and at $\pm\infty$ since $\chi (0)=1$ and
$\chi (\pm\infty )=0$. This can be better written as a row vector with
components $((e^{2i\pi\chi |_{{\bf R}^{-}}}-1),(e^{2i\pi\chi |_{{\bf R}^{+}}}-1))$ and the first component under the isomorphism $\psi ^{-}$ corresponds to
$(e^{-2i\pi\theta}-1)\in C_{0}({\bf R})$ and the second component under the
isomorphism  $\psi ^{+}$ corresponds to
$(e^{-2i\pi\theta}-1)\in C_{0}({\bf R})$. We can see the above
element $exp[p_{\pm}]$ as an element in
$(C_{0}({\bf R}^{-})\oplus C_{0}({\bf R}^{+}))\rtimes _{\phi}{\bf Z}_{2}\cong M_{2}(C_{0}({\bf R}))$ as the following $2\times 2$ matrix

$$\frac{1}{2}\sv{(e^{-2i\pi\theta}+1)\quad \pm e^{-2i\pi\theta}\mp }{\pm e^{-2i\pi\theta}\mp 1\quad e^{-2i\pi\theta}+1}$$
 Let us call this matrix $W$ and for simplicity denote it as
$W:=\sv{a\quad b}{c\quad d}$ Then the $+$ correspond to the projection $p_{+}$
under the isomorphism and similarly for the $-$ sign.\\

Now we incorporate the homotopies in order to see to what explicit elements
the projections $p_{\pm}$ correspond to. We know that there exists a homotopy
between

$\frac{1}{2}\sv{1\quad 1}{1\quad 1}$ and $\sv{1\quad 0}{0\quad 0}$\\
and between\\
$\frac{1}{2}\sv{1\quad -1}{-1\quad 1}$ and $\sv{0 \quad 0}{0\quad 1}$\\
given by\\
$u_{0}=\sv{1 \quad 0}{0\quad 1}$ and $u_{1}=\frac{1}{\sqrt{2}}\sv{1 \quad -1}{1\quad 1}$ with $u_{t}\frac{1}{2}\sv{1 \quad 1}{1\quad 0}u_{t}^{*}$\\
which is equal to\\
$\sv{1\quad 0}{0\quad 0}$ for $t=1$ and to $u_{0}$ for $t=0$.\\
This can be also written as\\

$$\sv{cos(t\frac{\pi}{4})\quad -sin(t\frac{\pi}{4})}{sin(t\frac{\pi}{4})\quad cos(t\frac{\pi}{4})}.$$

Hence

$$u_{1}Wu_{1}^{*}=\frac{1}{2}\frac{1}{2}\sv{1\quad 1}{-1\quad 1}
\sv{a\quad b}{b\quad a}\sv{1\quad -1}{1\quad 1}\sv{1\quad 1}{-1\quad 1}\sv{a-b\quad b+a}{b-a\quad a+b}=$$

$$=\sv{0\quad 2(a+b)}{2(b-1)\quad 0}$$

To the last matrix we substitute the values of $a$, $b$ from our earlier
computations of the matrix $W$ and get that

$$\sv{0\quad 2(a+b)}{2(b-1)\quad 0}=
\frac{1}{2}\sv{0\quad e^{-2i\pi\theta}(1\pm 1)+(1\mp 1)}{(\pm -1)e^{-2i\pi\theta}+(\mp 1-1)\quad 0}$$
 So we can see that

$[p_{+}]\mapsto [\sv{0 \quad e^{-2i\pi\theta}}{-1 \quad 0}]$

and

$[p_{-}]\mapsto [\sv{0 \quad -1}{-e^{-2i\pi\theta} \quad 0}]$

and we see that their images \emph{are homotopic}, namely $exp([p_{+}])$ is homotopic
to $exp([p_{-}])$ thus by this homotopy we lose one generator and so the generator of $K_{0}(C_{0}({\bf R})\rtimes _{\phi}{\bf Z}_{2})={\bf Z}$ is

$[\sv{0 \quad e^{-2i\pi\theta}}{-1 \quad 0}]$\\

We now turn our attention to the cyclic cohomology of the algebra $A:=C_{0}({\bf R})\rtimes _{\phi}{\bf Z}_{2}$ which is the corresponding $C^{*}$-algebra to
the Mobius foliation after having proved above that
$K_{0}(A)={\bf Z}=$. The cyclic cocycles are traces of $A$.
One usually looks at traces which are invariant under the holonomy action. In
our case we take a holonomy invariant transverse measure, namely a measure
$\mu $ on ${\bf R}$ which is the transversal such that $\mu\circ\phi =\mu$,
namely it is invariant under the action $\phi $ of the holonomy group
${\bf Z}_{2}$ on ${\bf R}$. In general an invariant measure has the property
that given any map $f:{\bf R}\rightarrow {\bf C}$ one has
$\int d\mu (x)f(x)=\int d\mu (x)f(-x)$. Such measures are very common indeed
(e.g. the Lebesgue measure has this property) but there are also many measures
on ${\bf R}$ which do not have this property. Having picked such a holonomy
invariant measure on the transversal ${\bf R}$ of the Mobius foliation we can
define a trace $\tau _{\mu}$ on $A$ (namely a closed cyclic 0-cocycle of $A$
 denoted $\tau _{\mu}\in HC^{1}(A))$ as follows:

$$\tau_{\mu}(F)=\int F(e)d\mu$$

where $F:{\bf Z}_{2}\rightarrow C_{0}({\bf R})$. (Aside: But we have to see
what is the transverse fundamental cyclic cocycle of the foliation used in
\cite{z} to define a numerical invariant for foliations; this transverse
fundamental cyclic cocycle has dimension equal to the codimension of the
foliation,
clearly the codim of the Mobius foliaiton is $1$. But we have to understand
the cyclic cohomology of the crossed product algebra $A$ first. Here we discuss
cyclic 0-cocycles because they appear in the gap labelling problem, the gap
labells come as pairings between K-classes and cyclic 0-cocycles).
 However this
is not a convenient choice since the pairing of $\tau_{\mu}$ with the generator
 $[p_{+}]$ (and hence any element since any element can be written as a
 multiple of the generator) of $K_{0}(A)$ vanishes.\\

We note here that ${\bf Z}$ (which is equal to $K_{0}(A)$) has only 2
generators, the one we have picked and minus that, but they both vanish when
paired with $\tau_{\mu}$.\\

We can do something elese though: since we have the map evaluation at the
fixed point $0$
$ev_{0}:A\rightarrow {\bf C}\rtimes _{\phi}{\bf Z}_{2}\cong C^{0}{\bf Z}_{2}$
where $C^{0}{\bf Z}_{2}$ denotes the group algebra of ${\bf Z}_{2}$,
instead of picking a holonomy invariant measure, we can pick a representation
$(\rho ,V)$ of the group algebra $C^{0}{\bf Z}_{2}$ onto a vector space $V$
with $\rho :C^{0}{\bf Z}_{2}\rightarrow End(V)$ and this in turn has a well
defined matrix-like trace $Tr:End(V)\rightarrow {\bf C}$. So the composition
$Tr\circ\rho\circ ev_{0}:A\rightarrow {\bf C}$ is also a trace. We pick for
example the following representation $\rho$ of ${\bf Z}_{2}$ onto ${\bf C}$:
$e\mapsto 1$ and $\epsilon\mapsto -1$. Using this we can easily see that
the pairing between K-classes and the cyclic 0-cocycle $\tau_{\rho}$ gives
$$<([p_{+}]),(\tau_{\rho})>=-1$$.\\

Now in \cite{z} we used the transverse fundamental
cyclic cocycle of the foliation in order to take pairings with K-classes. The
transverse fundamental cyclic cocycle is a cyclic $q$-cocycle where $q$ is the
codimension of the foliation and one needs the transverse bundle of the 
foliation to be \textsl{orientable} in order to be able to define it. For the 
Mobius foliation the codimension is 1 but the \emph{transverse bundle} is 
\emph{not orientable} because of the map $\phi$ which flips the sign and thus 
the transverse fundamental cyclic cocycle does 
not exist for the Mobius foliation. We need to find another example, either
the folaition of the annulus or the Reeb foliation of the 2-torus.

Understand restrictions that transverse bundle must be orientable and existence
 of a holonomy invariant transverse measure. The (partial) Bott connection 
$\theta$ is flat when restricted to the leaves but it still may have holonomy;
what is the relation between the holonomy of it and the holonomy $G_{x}^{x}$?
Are they the same? It seems to be so...
\\
\\
{\bf Acknowledgement:} We would like to thank Professor Johannes Kellendonk for his valuable help and guidance in
performing this K-Theoretic computation.

\end{document}